\documentclass[twocolumn,pra,superscriptaddress]{revtex4} 
\usepackage{epsfig}
\usepackage{graphicx}
\usepackage{amsmath}

\usepackage{color,colordvi,graphicx,pstcol}

\newcommand{\ket}[1]{\ensuremath{\left| #1 \right\rangle}}

\newcommand{\den}[2] {\ensuremath{{\rho}_{ #1 #2 }}}

\newcommand{\bnbar} {\ensuremath{b_{2\bar{k}}}}
\newcommand{\bDnbar} {\ensuremath{b^+_{2\bar{k}}}}

\newcommand{\cn} {\ensuremath{b_{3k}}}

\newcommand{\cDn} {\ensuremath{b^+_{3k}}}
\newcommand{\rhonn} {\ensuremath{\rho_{\bar{k}k}}}
\newcommand{\rhozz} {\ensuremath{\rho_{\bar{0}0}}}
\newcommand{\ak} {\ensuremath{a_{q2}}}
\newcommand{\Ak} {\ensuremath{A}}
\newcommand{\sr} {superradiance}

\newcommand{\beq}{\begin{equation}}
\newcommand{\eeq}{\end{equation}}

\begin{document}

\title{Theory for Raman superradiance in atomic gases}
\author{Tun Wang}
\affiliation{Department Of Physics, University of Connecticut,
Storrs, CT 06269}

\author{S. F. Yelin}
\affiliation{Department Of Physics, University of Connecticut,
Storrs, CT 06269} \affiliation{ITAMP, Harvard-Smithsonian Center
for Astrophysics, Cambridge, MA 02138}

\date{\today}

\begin{abstract}
A mean field theory for Raman superradiance (SR) with recoil is presented, where the typical SR signatures are recovered, such as quadratic dependence of the intensity on the number of atoms and inverse proportionality of the time scale to the number of atoms. A comparison with recent experiments and theories on Rayleigh SR and collective atomic recoil lasing (CARL) are included. The role of recoil is shown to be in the decay of atomic coherence and breaking of the symmetry of the SR end-fire modes.
\end{abstract}

\maketitle

\section{Introduction}
\label{s_introduction} Superradiance, first proposed by
Dicke~\cite{DickeSuper}, is the enhanced radiation from a collection of coherently decaying dipoles. It has been studied extensively
theoretically (see review~\cite{SuperRev} and references therein) and
has been observed in many different systems, including thermal
gases~\cite{SuperRev}, 
and
Bose-Einstein condensates (BECs)~\cite{SuperBEC,RamanAmpMatter,SuperRamanBEC,ThermalSuper}. 
There is mostly agreement now on the fact that the collectivity is responsible for superradiance, which is the same no matter whether the medium consists of Bosons or
Fermions~\cite{atomicFWMfermionsVsBosons,MatterAmpFermions,CooperationJJ}.
In the case of BEC, collectivity can be observed as matter wave
stimulation, or ``Bosonic
enhancement''~\cite{atomicFWMfermionsVsBosons}. BEC is unique in
that there is negligible Doppler broadening and the recoil
momentum is measured easily and in fact was recently used to demonstrate
BEC superradiance
~\cite{SuperBEC,RamanAmpMatter,SuperRamanBEC,ThermalSuper}. In
particular, superradiance can be described by ``collective
atomic recoil laser'' (CARL) equations in the bad cavity
regime~\cite{GainBunching,propagationCARL,SuperBEC}. Collective gain can be observed with CARL in the sense that it depends nonlinearly on the
density~\cite{ObserveLasingCARL} and thus does not occur for atomic
densities below a certain critical value ~\cite{RIRcompCARL}. 

Most experiments on superradiance were done using pulsed pump
lasers to ``instantaneously'' invert a two-level system. The quantum
stage of superradiance, where the radiation field builds up from
vacuum fluctuations, can then be modeled to start only after the pump laser is turned
off~\cite{SuperRev}. For this case, pump lasers obviously have to be strong; at the same time, experiments done with BECs use only weak pump fields. We therefore call the one {\em strong pump superradiance} and the other {\em
weak pump superradiance}. In the latter case, the quantum stage happens
while the pump field is still on. In this article, we will focus on weak pump superradiance. Note that in this case the maximum instantaneous superradiance rate is limited by the
pump laser intensity, while for strong pump superradiance no such
limitation exists.

Mostly, earlier research concentrated on so-called {\em Rayleigh}
superradiance~\cite{TheorySuperBEC,TheorySuperBEC1,SuperBECoc,OnsetAmp,WaveMixingOpticalBEC,RayleighSemiAndQuantum}
which happens for transitions between different center of mass
(c.m.) states while the internal state remains
unchanged~\cite{AtomDiffCARL}. We will here discuss {\em Raman}
superradiance, where there are two different internal ground
states for the pump and the superradiant transition. Recoil and
different c.m. states are taken into account here as well, but
are, as we will show, of lesser consequence. It turns out that
Raman superradiance otherwise follows the same basic patterns as
Rayleigh superradiance.
Although superradiance with Raman pumping has been analyzed in Ref.~\cite{CoThreeLevel}, 
the recoil effect was ignored and the Raman pumping time was
assumed to be short compared with the superradiance time. It will
be shown in this paper that recoil induces the decay of Raman
coherence and may make the superradiant modes asymmetric. In
Ref.~\cite{SuperInCW} an incoherent cw pump laser was considered
numerically, also leading to superradiance. Recently, M. M. Cola,
{\em et al.}~\cite{RamanSuperBEC} presented a quantum theory to
describe the Raman superradiance experiments with
BECs~\cite{RamanAmpMatter,SuperRamanBEC,ThermalSuper}. In
comparison, our analysis can be applied to both thermal atoms and
BECs with emphasis on the effect of recoil. We also discuss the
connection with CARL using stability analysis. In addition, we
consider the asymmetry of superradiant modes as the pump laser
setup is changed which helps to understand the underlying physics
of superradiance.

This paper is organized as follows: In Sec.~\ref{s_formalism} we
derive the dynamical equations to describe Raman superradiance.
These equations are used to analyze the stability conditions in
Sec.~\ref{s_instability}. Numerical calculations in comparison
with experiments are included in Sec.~\ref{s_numerics}. Discussion
and conclusion follow in Sec.~\ref{s_conclusion}.
\section{Model}
\label{s_formalism} 
\begin{figure}[ht]                                                           \begin{picture}(200,100)(10,20)                                    
\put (90,105){\ket1} \put (30,20){\ket3} \put(150,20){\ket2}   
\multiput(70,90)(10,0){5}{\line(1,0){7}}                                        
\put(70,100){\line(1,0){50}}                                                    
\put(10,30){\line(1,0){50}}                                                     
\thicklines \put(130,30){\line(1,0){50}} \thinlines                             
\put(155,30){\vector(-3,4){45}}                                                 
\put(155,30){\vector(-3,4){45}}                                                 
\put(80,90){\vector(-3,-4){40}} \put(100,90){\vector(3,-4){40}}                 
\put(35,60){$a_{\vec{q}3}$} \put(105,60){$a_{\vec{q}2}$}                        
\put(140,60){$\Omega$} \put(90,92){$\Delta$}                                    
\end{picture}                                                                   
\caption{\protect\label{fig:levels} Center of mass manifolds                     associated with three-internal-state atomic system. State $\ket{2}$ is the one-particle state of the initial BEC. Both Rayleigh      
transition and Raman transition are present with Raman  
field $a_{\vec{q}2}$ and Rayleigh field $a_{\vec{q}3}$. The pump  
laser Rabi frequency $\Omega$  is much smaller
than the detuning $|\Delta|$.}  
\end{figure}
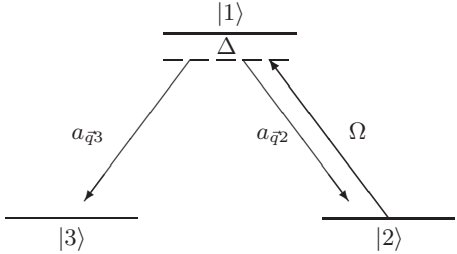

We consider a three-level $\Lambda$-type
atomic system with excited state \ket1 and two ground states \ket2
and \ket3 (Fig.~\ref{fig:levels}). When the detuning of a pump
laser is much larger than both, its Rabi frequency and the maximum
Rabi frequency of the superradiant field, the interaction picture
Hamiltonian of this system under dipole and rotating-wave
approximation
reads~\cite{PhaseDepSpectrum,PhaseBetweenBEC,MeasurePhaseBEC}
\begin{eqnarray}
    H & = &\Psi^+_2H_{cm}\Psi_2+\Psi^+_3(H_{cm}-\hbar
    \delta_3)\Psi_3+H_f \label{eq:FarDetuningH0} \nonumber\\
      &   & +\sum_{\vec{q}}\hbar g^*_{3,\vec{q}}e^{-i(\vec{q}-\vec{k}_0)\cdot \vec{r}}\Psi^+_3a^+_{\vec{q}3}\Psi_3+H.c. \label{eq:FarDetuningRayleigh} \nonumber\\
      &   & +\sum_{\vec{q}}\hbar g^*_{2,\vec{q}}e^{-i(\vec{q}-\vec{k}_0)\cdot \vec{r}}\Psi^+_2a^+_{\vec{q}2}\Psi_3+H.c.\;, \label{eq:FarDetuningRaman}
\end{eqnarray}
with coupling constants $g^*_{2,\vec k}=i\sqrt{\frac{\hbar
kc}{2\epsilon_0V}}\hat{\epsilon_2}\cdot\vec{d}_{12}\frac{\Omega^*}{\Delta}$
and $g^*_{3,\vec k}=i\sqrt{\frac{\hbar
kc}{2\epsilon_0V}}\hat{\epsilon_3}\cdot\vec{d}_{13}\frac{\Omega^*}{\Delta}$, in what follows assumed to be real. $\hat{\epsilon_i}$ is the polarization direction.
While the c.m. Hamiltonian is
$H_{cm}=-\frac{\hbar^2}{2m}\nabla^2$ with $m$ being the mass and
$\hbar$ being the Planck constant, the Hamiltonian of the optical
fields is $H_f=\sum_{\vec q} \hbar q\,c\, a^+_{\vec q2}a_{\vec
q2}+\hbar q\,c\,a^+_{\vec q3}a_{\vec q3}$, where $a_{\vec q2}$
($a_{\vec q3}$) is the field annihilation operator for the
transition between \ket1 and \ket2 (\ket3), and $\vec q$ the
momentum of the radiation field. $\Psi_i$ is the atomic field
operator, $V$ the quantization volume, $\Omega$ and $\Delta$ the
pump field Rabi frequency and detuning, $\delta_3$ the ac-Stark
shift due to the pump laser, $d_{12}$ ($d_{13}$) the dipole moment
between \ket1 and \ket2 (\ket1 and \ket3). In
Eq.~(\ref{eq:FarDetuningRaman}), the second line describes the
Rayleigh transition, the third line the Raman transition. The
ratio of $g_{2,\vec q}/g_{3,\vec q}$ determines the branching
ratio between Rayleigh and Raman
superradiance~\cite{RamanAmpMatter,SuperRamanBEC}. While the
Rayleigh superradiance has been studied extensively
\cite{TheorySuperBEC,TheorySuperBEC1,TheorySuperBEC2}, this paper
will focus on Raman superradiance.

Using Fock representation, Eq.~(\ref{eq:FarDetuningH0}) can be written as
\begin{equation}
    H/\hbar=\sum_{j,k}
    \omega_{k}b^+_{jk}b_{jk}+(g_{2}\sum_{q,k}b^+_{2\bar{k}}a^+_{2q}b_{3k}+h.c.)+\sum_q\omega_qa^+_{q2}a_{q2}
    \label{eq:FockH}
\end{equation}
where $b_{jk}$ ($j=2,3$) annihilates an atom in state $\ket{j}$
with momentum k and energy $\omega_k=\hbar^2 k^2/2m$.
$b_{2\bar{k}}\equiv b_{2,k+k_0-q}$,
$\omega_q=cq-\omega_0+\omega_{23}+\delta_3$, $\omega_0=ck_0$ and
$\omega_{23}$ is the atomic energy difference between \ket2 and \ket3.
For simplicity, the vector arrows from  $\vec{q}$ and $\vec{k}$ have been
dropped here. We assume that $g_{2,q}\approx g_2$ only weakly depends on 
$q$ for the relevant range of modes. From this form it is obvious that the total population  on \ket2
and \ket3 is conserved. The matter wave mode $b_{3k}$ is
coupled to different \bnbar\ for different optical modes $q$. When
the detuning is large, however, collective linewidth or multiple scattering
can be neglected~\cite{LightScattBoseFermi} and we can drop the
coupling between different modes. In this article, we also neglect the depletion of BEC due to
other modes. Raman transitions in
different directions can thus be considered independently. From
Eq.~(\ref{eq:FockH}), Maxwell-Bloch equations can be derived:
\begin{subequations}
\label{eq:QAve}
\begin{eqnarray}
    \frac{d}{dt}A & = & -i \omega_k \Ak-i g_2 \sum_k \rhonn-\kappa \Ak,\label{eq:QAveField}\\
    \frac{d}{dt}\rhonn & = & -i(\omega_{k}-\omega_{\bar{k}})\rhonn-i
    g_2(1-2N_k)\Ak,\label{eq:QuantumAveRaman}\\
    \frac{d}{dt}N_k & = & i(g_2\rhonn \Ak-c.c.) \;, \label{eq:QAver33}
\end{eqnarray}
\end{subequations}
where $\Ak=\langle\ak\rangle$, $\rhonn=\langle\bDnbar\cn\rangle$,
$N_k=\langle\cDn\cn\rangle$, with $N_k$ being the number of atoms
in state $|3k\rangle$. $\kappa$ is the effective radiation field decay rate,
if we neglect propagation in the mean field approximation~\cite{CoSuperfluorescence,TheorySuperBEC2}. 
This approximation works well when the medium is optically thin at
the pump frequency, which is the case here since the pump field is far detuned from resonance. 
With $L$ and $D$ the length and diameter of the medium and $\lambda$ the wavelength of the superradiant transition, the Fresnel number $F=\frac{D^2}{L\lambda}$ gives approximately the number of modes that fit in the medium in axial direction. If it is around or bigger
than 1 as in the experiments~\cite{RamanAmpMatter,SuperRamanBEC},
then $\kappa=c/2L$ for axial modes (also called ``end-fire
modes"~\cite{SuperBEC}), which are the modes having largest gain
for superradiance, and
$\kappa_{\rm off}\geq\frac{c}{2L}(\frac{1}{F}+1)$ for
off-axial modes~\cite{CoSuperfluorescence}. It
will be shown in Sec.~\ref{s_instability} that in
experiments~\cite{RamanAmpMatter,SuperRamanBEC}, $\kappa$
dominates over all the other relevant characteristic rates and therefore
makes the end-fire modes most likely to superradiate. In the
following, we assume all superradiant modes to be axial.

In a BEC, only the $k=0$ state is present, and thus
Eqs.~(\ref{eq:QAve})
become
\begin{subequations} 
\label{eq:BEC}
\begin{eqnarray}
    \frac{d}{dt}A & = & -i \omega_k \Ak-i g_2 N \rhozz-\kappa \Ak,\label{eq:BECField}\\
    \frac{d}{dt}\rhozz & = & i\,\omega_r\rhozz-i
    g_2(1-2N_0)\Ak \label{eq:BECRaman}\\
    \frac{d}{dt}N_0 & = & i(g_2\rhozz \Ak-c.c.) \label{eq:BECr33}
\end{eqnarray}
\end{subequations}
where $\rhozz=\langle b^+_{2,\bar{0}} b_{3,0}\rangle$ with
$b_{2,\bar{0}}=b_{2,k_0-q}$, the recoil energy
$\hbar\omega_r=\hbar^2 (k_0-q)^2/2m$, and $N$ the total number
of atoms in the system. Because we assume $\kappa$ to be very
large it follows from Eq.~(\ref{eq:BECField}) that
\begin{equation}
    \Ak\approx -ig_2N\rhozz/\kappa \label{eq:AEV}
\end{equation}
Substituting \Ak\, into Eqs.~(\ref{eq:BECRaman},\ref{eq:BECr33}), we arrive at 
\begin{eqnarray}
    \frac{d}{d\tau}\rhozz & = &i\omega_r\rhozz
        -g^2_2(2N_0-1)\rhozz \label{eq:SuperMean}\\
    \frac{d}{d\tau}N_0 & = & -2g^2_2|\rhozz|^2\;. \nonumber
\end{eqnarray}
Here, we scale the time such that $\tau=N t$. It is therefore obvious that the timing of the resulting process scales with $1/N$, in the same way as in traditional superradiance \cite{SuperRev}. From Eq.~(\ref{eq:AEV}) we know that
the output field amplitude \Ak\, is proportional to $N$ and thus the
intensity is proportional to $N^2$. These are typical
characteristics of \sr. Note that without recoil
$\omega_r=0$, Eqs.~(\ref{eq:SuperMean}) are completely equivalent
to Eqs. (6.36) of~\cite{SuperRev}, which describe {\em standard
superradiance}: a radiation cascade down the pseudo-spin ladder from
$J_z=N/2$ to
$J_z=-N/2$,
giving a hyperbolic secant solution for the dependence of the upper-level population on time~\cite{SuperRev}. For BEC, the term with
$\omega_r$, which is due to recoil, only contributes to the
phase evolution of Raman coherence, not to its decay, while for thermal atoms recoil does induce the decay of Raman coherence, as discussed in the next
paragraph.

For thermal atoms, Eq.~(\ref{eq:QuantumAveRaman}) describes
quantum diffusion as well as generation of Raman coherence . In
particular, the term $-i(\omega_{k}-\omega_{\bar{k}})\rhonn$ in
Eq.~(\ref{eq:QuantumAveRaman}) shows that coherence stored in
different levels experiences quantum diffusion, since the term
will have different values for different $k$. To understand how
the quantum diffusion works, we assume Raman coherence has been
generated uniformly for all levels, which means
$\rhonn(0)=\rho(0)\,p_k$ with $\rho(0)$ being the coherence for
one level and $p_k$ the probability distribution of atom at level
$k$. If we set field amplitude $\Ak$ to zero, the solution of
Eq.~(\ref{eq:QuantumAveRaman}) is
$\rhonn(t)=\rho(0)p_k\,e^{-i(\omega_{k}-\omega_{\bar{k}})t}$. The
coherence is then
$\rho(t)=\sum_k\rhonn(t)=\rho(0)\sum_kp_k\,e^{-i(\omega_{k}-\omega_{\bar{k}})t}$.
To proceed, we need to specify $p_k$ at temperature $T$. Here,
either Bose-Einstein distribution for Bosons or Fermi-Dirac
distribution for Fermions are appropriate. For simplicity,
however, we assume Lorentzian distribution
$p_k=\frac{1}{\pi}\frac{\delta p}{k^2+\delta p^2}$ with $\delta
p^2/2m=k_B T/2$, which describes the atoms well even at sub-recoil
temperatures~\cite{SpatialCorrelation}.
The summation can be approximated by an integral and it follows
that the Raman coherence decays exponentially

\begin{equation}
\rho(t)=\int \frac{1}{\pi}\frac{\delta p}{k^2+\delta p^2}
\rhonn(t) dk= \rho(0)e^{i \omega_\Gamma t}e^{-\Gamma\,\,t}
\end{equation}
where $\Gamma=2k_0\sqrt{k_BT/m}\sin{\theta/2}$,
$\omega_\Gamma=2\hbar k^2_0\sin^2{\theta/2}$ and $\theta$ is the
angle between $\vec{q}$ and $\vec{k}_0$. It is clear now that the
decay rate $\Gamma$ depends on the pump laser direction
$\hat{k}_0$ relative to the
superradiant pulse direction $\hat{q}$. 
Thus, Eq.~(\ref{eq:QuantumAveRaman}) can be rewritten as
\begin{equation}
        \frac{d}{dt}\rho  =  (i\omega_\Gamma-\Gamma)\rho-i
    g_2(1-2\den33)\Ak \; , \label{eq:QAveRaman}
\end{equation}
where $\den33=\sum_kN_k$ is the population in state \ket3. If we
would use a Gaussian rather than Lorentzian density of states, the inverse
$1/e$ decay time would be  $\sqrt{2}\,\Gamma$ rather than $\Gamma$. As an example, for
Rb at the Doppler limit temperature of $143\mu K$,
$\Gamma=1.35\times10^6 s^{-1}$.

Comparing the result for thermal atoms in Eq.~(\ref{eq:QAveRaman}) with the result for a BEC in Eq.~(\ref{eq:BECRaman}),
we see that thermal distribution contributes additional coherence decay,
otherwise these equations are the same as expected. We can therefore generalize the results to
\begin{subequations} 
\label{eq:Raman}
\begin{eqnarray}
    \frac{d}{dt}A & = & -i \omega_k \Ak-i g_2 N \rho-\kappa \Ak, \\
    \frac{d}{dt}\rho & = & i\,\omega_\Gamma\rho-i
    g_2(1-2\den33)\Ak-\kappa_R\,\, \rho \\
    \frac{d}{dt}\den33 & = & i(g_2\rho \Ak-c.c.)\;,
\end{eqnarray}
\end{subequations}
where the total coherence decay $\kappa_R=\kappa_R'+\Gamma$. $\kappa_R'$ can be introduced phenomenologically to contain collisions, magnetic gradients,
etc., and $\Gamma=0$ for BECs. These equations are now analogous
to Eqs.~(13-15) in Ref.~\cite{RamanSuperBEC}, but can be applied
to both, BEC and thermal atoms.

\section{Linear stability analysis}
\label{s_instability} In this section, we will determine
the necessary conditions for Raman superradiance to happen, which is easiest using linear stability analysis~\cite{BosonicAmpAtomPairs,AtomDiffCARL,DopplerCARL}. Obviously, $A=0$, $\rho=0$, and $\rho_{33}=1$ give a stationary solution of Eqs.~(\ref{eq:Raman}). Rewriting Eqs.~(\ref{eq:Raman}) for $A=0+\delta A$, $\rho=0+\delta\rho$, and $\rho_{33}=1+\delta\rho_{33}$ leads to a two-dimensional linear system with the characteristic equation
\begin{eqnarray}
    S^2+\left(i\left(\omega_k-\omega_\Gamma\right)+\kappa+\kappa_R\right)S
    +\left(-i\omega_\Gamma+\kappa_R\right)\kappa \nonumber \\
    -Ng^2_2+\omega_k\omega_\Gamma+i\omega_k\kappa_R=0\;.
    \label{eq:instability}
\end{eqnarray}
(The third equation is equivalent to zero in this case and can be dropped.)
In comparison with the cubic instability equation for Rayleigh
superradiance~\cite{AtomDiffCARL}, this is a quadratic equation.
The physical reason for such a change is that for a Rayleigh transition, the
initial and final internal states are the same and thus only atoms
with different c.m. states may contribute to the gain (see Eq. (49)
of Ref.~\cite{AtomDiffCARL}); for a Raman transition, the
initial and final internal states are different and thus all atoms
contribute to the gain regardless of the c.m. states.

The above
quadratic equation has two roots for $S$, $S_+$ and $S_-$. Since $S$ is the exponent of the state vector [$(\delta A, \delta\rho) = (\delta A(0),\delta\rho(0)) \exp S t$], the zero solution becomes unstable if at least one of $S_+$ or $S_-$ has a positive real part. The larger real part (let's call the respective root $S_+=S_+'+iS_+''$) is therefore defined as {\em instability factor}. If $S_+'>0$,
then the system is dynamically unstable, from which the threshold pump
intensity can be derived. From Eq.~(\ref{eq:instability}), it can
be easily seen that $S_+'$ depends nonlinearly on the number of atoms $N$. Note that
nonlinear dependence on $N$ is the essence of {\em collective
instability}~\cite{AtomDiffCARL}. In the bad cavity regime as is the case for the
experiments~\cite{RamanAmpMatter,SuperRamanBEC}, $\kappa$ is large, the system therefore depends linearly on the atomic density and therefore may display superradiant behavior. In a good
cavity, however, $\kappa$ is much smaller and thus the
collective gain depends on the density
nonlinearly~\cite{propagationCARL,ObserveLasingCARL}.

Since $\omega_\Gamma$ and $\omega_k$ can be shown to have only a minor
effect on $S_+'$ under the experimental
conditions of Refs.~\cite{RamanAmpMatter,SuperRamanBEC}, we set
$\omega_\Gamma=\omega_k=0$. In this case, the instability factor
is
\begin{equation}
    S_+'=\frac{-(\kappa+\kappa_R)+\sqrt{(\kappa+\kappa_R)^2+4(Ng^2_2-\kappa\kappa_R)}}{2}
    \label{eq:DegSol}
\end{equation}
In particular, for vanishing coherence decay $\kappa_R=0$, the system is unstable and therefore superradiant for any pump laser power. This is different from the case of Rayleigh
superradiance which always has a non-zero threshold pump laser intensity
\cite{AtomDiffCARL}. In the case of thermal atoms, however, $\kappa_R$ can be considerable, and the threshold pump intensity is quite high in the bad cavity limit. This explains why collective gain was not observed in Refs. \cite{RamanAmpMatter,SuperRamanBEC}. It should be possible experimentally to minimize the decay due to quantum diffusion if the pump laser is collinear with the sample. Raman superradiance or collective gain might perhaps be observed in this case even in thermal atoms.

When $\kappa$ is much larger than any other frequency in
Eq.~(\ref{eq:instability}), i.e., $\kappa\gg
\sqrt{Ng^2_2}$, $\omega_k$, $\omega_r$, as in the experiments~\cite{RamanAmpMatter,SuperRamanBEC}, the instability factor can be simplified to
\begin{equation}
    S_+'\simeq\frac{1}{\kappa}\left\{Ng^2_2-\kappa_R\kappa\right\}
    \label{eq:DegSol1}
\end{equation}
In this case, $S_+'$ is linear in $N$, which
means experiments in Ref.~\cite{RamanAmpMatter,SuperRamanBEC} would be purely in the superradiant regime.

\section{Numerical simulations} \label{s_numerics}
To compare our theory with experiments, we solve Eqs.~(\ref{eq:Raman}) 
for both BEC and thermal atoms. In the simulations, we use the
initial value of $\rho(0)=(2/N)^{1/2}$, which is determined by
quantum noise~\cite{CoSuperfluorescence,StatSuper}. Other
parameters are calculated using the data in
Ref.~\cite{SuperRamanBEC}: $g_2= 0.5\times 10^6$ s$^{-1}$,
$\kappa= 1.76\times 10^{12}$ s$^{-1}$. $\omega_r$ is negligible in
this context.

\begin{figure}[ht]
    \centerline{\includegraphics[clip,width=.95\linewidth]{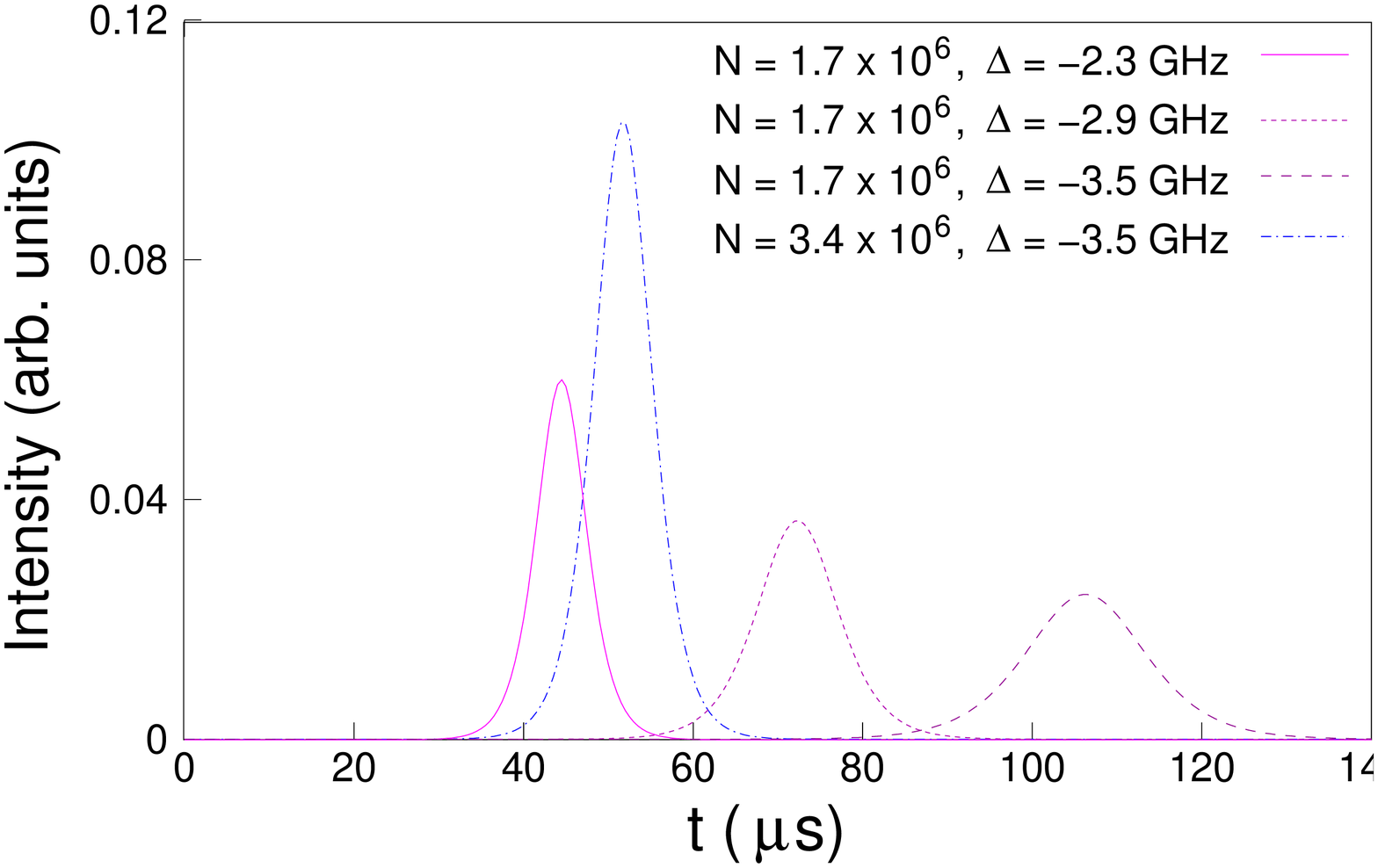}}
\centerline{\includegraphics[clip,width=.95\linewidth]{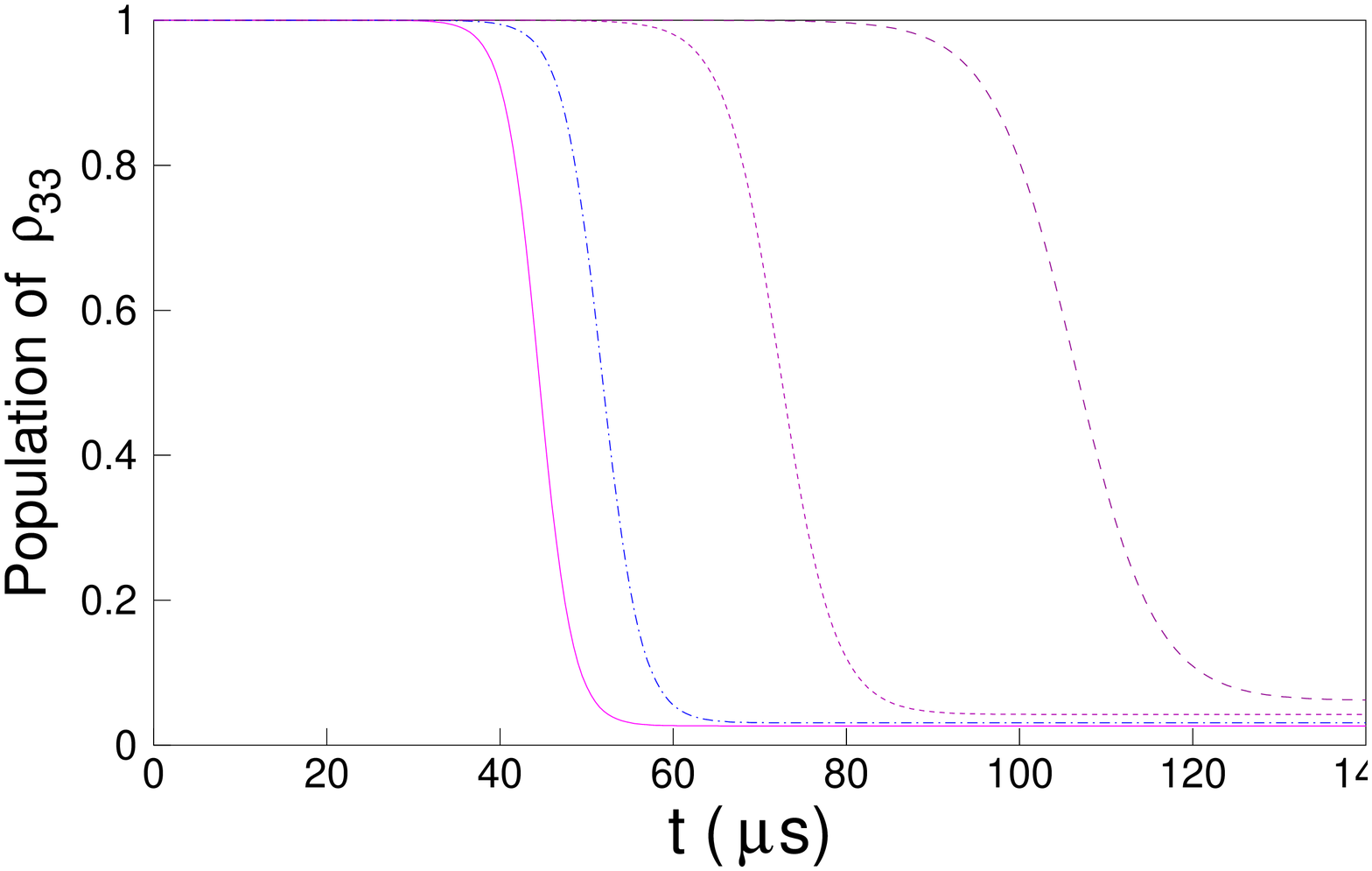}}
    \caption{\protect\label{fig:Kappa}Effect of number of atoms and detuning on the evolution of (a) the intensity $|A|^2$ and (b) the population $\den33$ as a function of time. Parameters used in the calculations are from Ref.~\cite{SuperRamanBEC}. The $1/N$ dependence of the delay time and the $N^2$ dependence of the maximum intensity can be clearly seen. The finite population left in state \ket3 is due to the decay of Raman coherence.}
\end{figure}

In Fig.~\ref{fig:Kappa} we show that the intensity of superradiance
is proportional to $N^2$ and the superradiance delay time is
proportional to $1/N$ at least as long as there is no Raman coherence decay, i.e., if we assume a BEC. The
numerical delay time $75-150\mu S$ also reproduces well the
experimental data~\cite{SuperRamanBEC,RamanAmpMatter}. 
%

\begin{figure}
\centerline{\includegraphics[clip,width=.95\linewidth]{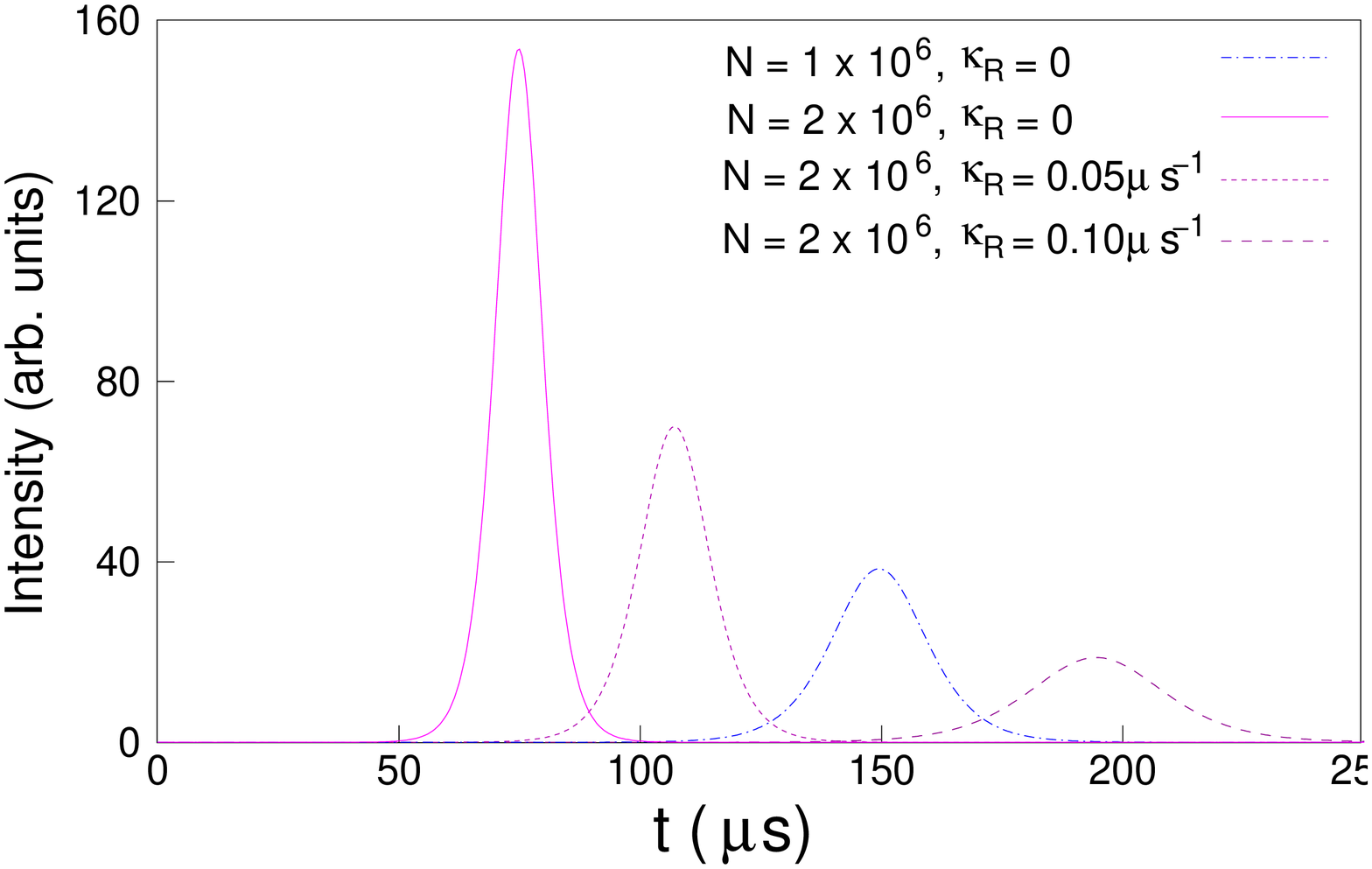}}

\centerline{\includegraphics[clip,width=.95\linewidth]{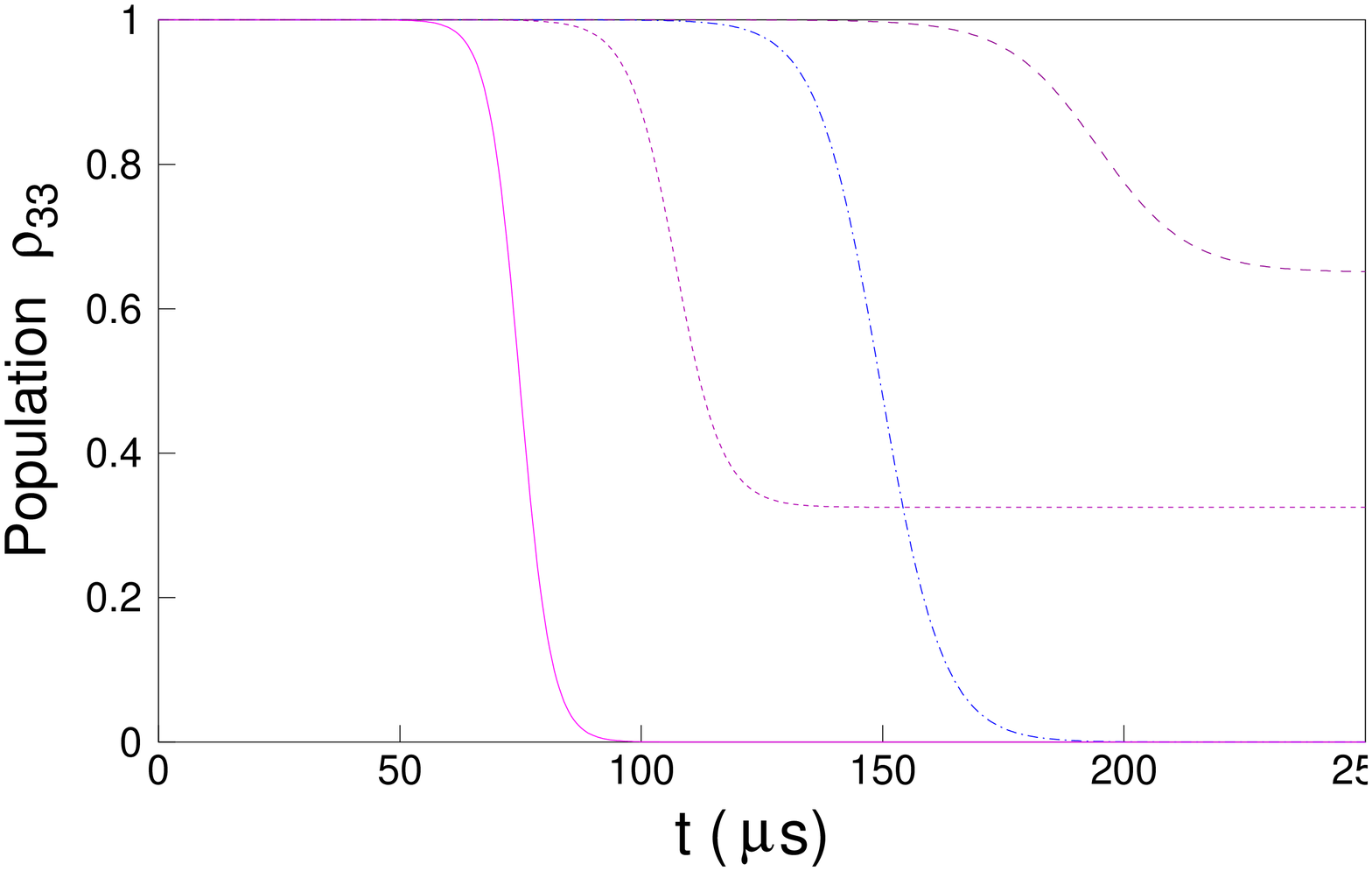}} %
    \caption{\protect\label{fig:KappaR} Effect of Raman coherence decay rate $\kappa_R$ on the evolution of (a) the intensity $|A|^2$ and (b) the population $\den33$ as a function of time. $\kappa_R>0$ is responsible for a longer superradiance delay time and a lower maximum intensity. For this figure, $N=2\times10^6$. Other parameters are the same as in Fig.~\ref{fig:Kappa}.}
\end{figure}

%
Figure~\ref{fig:KappaR} shows that when the Raman coherence decay rate
$\kappa_R$ is increased, the radiation intensity decreases and the
delay time increases. This is similar to two-level superradiance: dipole-dipole interaction decreases the coherence between atoms and thus competes with
superradiance. Because of the effective population mixing caused by Raman coherence decay there is always a finite number of atoms in the Rayleigh lower state \ket3 at any time for a finite $\kappa_R$. For thermal atoms at Doppler cooling
limit $T=143\mu K$, $\Gamma=1.35\times10^6 s^{-1}$ and the instability
factor $S_+'$ is smaller than zero and no superradiance happens.

The roles of photon and atomic coherence are intertwined for
superradiance. Collectivity can be attributed to either photons or
atoms, or both. In the case of weak pump superradiance, the pulse
exits the medium and thus decays much faster than the (atomic)
Raman coherence. Thus the intensity of the superradiant pulses is
small, and stimulation of photons by photons is not critical in
this case. For example, if in the calculation the pump laser is
turned off before all the atoms have radiated and then turned on
again, superradiance continues nearly at the same point it was
interrupted. This is true for an interruption that lasts longer
than the photon coherence time (which is here just the escape time
of the photons of about 1 ps), but shorter than the Raman
coherence time, which is between infinity and 1 ms in our
simulations. The conclusion is that atomic coherence is more
important than stimulated emission in this case
 for
superradiance to happen.

It was claimed in
Ref.~\cite{SuperRamanBEC} that the output photon number,
$N_p$, enhances the superradiance $\dot{N}_r\propto (N_r+N_p+1)$, where $N_r$
is the number of atoms having superradiated. 
Indeed, it was assumed that $N_p=N_r$~\cite{SuperRamanBEC}, then
$\dot{N}_r\propto (2N_r+1)$. However, no cavity was used
in~\cite{SuperRamanBEC}, which means the average $N_p$ is small
within the sample and can be neglected, as is done in
Ref.~\cite{SuperBEC,RamanAmpMatter}. Note that the collecting of
photons in the
(ring) cavity modifies the rate~\cite{TheoryCARL,SuperShortPulse}. 
But in a high Q cavity, the coupling between atoms and
field is strong and a perturbation approach of Fermi's golden rule
as used in Ref.~\cite{SuperRamanBEC} may not apply. Detailed
analysis of this is beyond this paper.


Now we consider the effect of the field decay rate $\kappa$ on
determining the direction of the superradiant field modes. If the Fresnel number $F$ is around $1$ as in
the experiments \cite{SuperRamanBEC,RamanAmpMatter}, the decay rate of the off-axial
modes is much bigger than that of the axial mode, and thus only the axial mode superradiates.
Since the decay rates of the Raman coherence for the axial modes
and their directly neighboring modes are almost equal to each other, the field
decay rate determines the radiation direction. On the other
hand, if $F$ is much bigger than $1$, the off-axial modes do not
have a decay rate much different from the axial ones, and thus they may also superradiate.
In this case, the quantum fluctuation stage
determines which modes are fired. The random dots in the
simulation of Ref.~\cite{TheorySuperBEC} show the effect of the fluctuations in this case.  In general, many modes might fire
simultaneously as long as the population in state \ket3 is not depleted. If the
delay time of one mode is shorter than the sum of the delay and superradiance time of the axial mode, then this mode
also superradiates. The same is obviously is also true for the competition
between Raman and Rayleigh superradiance~\cite{RamanAmpMatter}.

Let us consider the symmetry of the superradiating modes. The two
axial modes in opposite directions in the
experiments~\cite{SuperBEC,RamanAmpMatter,SuperRamanBEC} show
identical behavior: the recoil pattern is symmetric. One of the
reasons for this is that the field decay rate for these modes is
the same. Also the recoil induced decay
$\kappa_R$ is zero for a BEC. This is also true for an
initially fully inverted two level
system~\cite{SuperRev}.
However, if the pump laser is parallel to the sample axis (longitudinal
pumping), the recoil
induced decay for thermal atoms
can be cancelled if the superradiance mode is parallel to the pump
laser. This breaks the symmetry of the two axial modes and privileges the
parallel mode over the antiparallel one.
In Fig.~\ref{fig:KappaR} we see that the mode with small $\kappa_R$
is stronger than other modes and may suppress superradiance for them by depleting the population $\rho_{33}$. The broken
symmetry indicates that the equivalence of a three-level system with a far detuned pump laser and a two-level system does not hold in
this case. Note that if the atoms are not fully inverted, the
symmetry could  also be broken due to stored
coherence~\cite{Super1735}. However, for BECs, since recoil does
not contribute to the decay of Raman coherence significantly, two
superradiant modes would still fire even with longitudinal pump.

\section{Discussion and Conclusion} \label{s_conclusion}
Rayleigh superradiance does not happen without recoil. In
comparison to this, recoil is not critical for Raman superradiance to happen,
which means that atomic bunching and density grating pictures do not
apply for explaining Raman superradiance, as they do for Rayleigh
superradiance. Interference between pump laser and superradiance
output~\cite{TheorySuperBEC2} equally does not apply in a case where both transitions radiate light with different polarization. We therefore believe that collective
effects, which might be called Bosonic stimulation in the case of
Bosons, are the main players in Raman superradiance.

Interesting is the  relationship between Rayleigh
 and Raman superradiance. States with different
momentum may be considered to be orthogonal~\cite{TheorySuperBEC}
 in the same way as different internal levels, and thus the Rayleigh transition can be looked upon as a Raman transition between {\em different} motional
states~\cite{RIR92,DopplerCARL,RIR94}. Indeed, the gain
coefficients have a similar functional dependence on the atomic
density~\cite{RIRcompCARL}. In particular, in the case of thermal atoms with a pump laser not parallel to the sample axis, i.e., with large $\kappa_R$,
it can be shown from Eq.~(\ref{eq:DegSol}) that the instability factor $S_+'$ depends
linearly on $N$. This was the regime discussed in
~\cite{RIR92,DopplerCARL,RIR94} in which the Raman transition is
considered to be in the (linear) single-atom
gain regime~\cite{AmpLightAtomInBEC}. Although atom statistics are not
critical for superradiance~\cite{TheorySuperBEC2}, the Fermi momentum
$k_F$ in Rayleigh scattering is replaced by the relative momentum
difference in Raman scattering, thus the problem with a very short coherence time in the case of fermions due to recoil might be
overcome~\cite{MatterAmpFermions}. As is done
for Rayleigh superradiance~\cite{AtomInteractSuper}, also atom-atom
interaction can be
included, and will be presented in a forthcoming publication.

Finally, we would like to differentiate two concepts: collectivity and
collective gain or collective instability. Collectivity means that
all atoms in the system contribute to the same mode~\cite{DSP1},
while collective gain or collective
instability~\cite{ObserveLasingCARL} means that the gain depends on the number of atoms $N$ nonlinearly. While the experiments are
in the non-collective gain regime, collectivity still plays a major role in
Raman superradiance. Raman superradiance therefore shows that it is the
collective effect rather than ``Bosonic stimulation'' that is responsible for
superradiance~\cite{CooperationJJ,atomicFWMfermionsVsBosons,MatterAmpFermions}.
It was claimed~\cite{RIRcompCARL} that if the pump laser makes the two-photon detuning for superradiant mode zero, and thus the Rayleigh
transition corresponds to a Raman transition between different c.m.
states there would be a single-atom gain instead of
collective gain ~\cite{AmpLightAtomInBEC}. However, we tried to show that
even in a pure Raman transition, collective gain is still possible
if a cavity is included.

To conclude, we developed a mean field theory for Raman
superradiance. Raman superradiance does not necessarily have an intrinsic threshold for pump laser intensity
even if the decay of the optical field is included. We found that
recoil induced decay of Raman coherence may break the symmetry of
the two axial modes if the atoms are pumped longitudinally, in which
case it is possible to realize Raman
superradiance even in thermal atoms while at the same time it might not be possible to realize Rayleigh
superradiance. We also note that both the Rayleigh and Raman
superradiance experiments were done in the regime where the pump laser is far detuned, such as not to populate the excited state. What happens in the case of a resonant pump laser is under investigation presently.

\centerline{ACKNOWLEDGEMENTS} The authors gratefully acknowledge
useful and stimulating discussions with W. Ketterle, J.
Javanainen, M. Ko\v{s}trun and the support from NSF and the Research
Corporation.

\bibliography{RamanSuper}
\end{document}